\documentclass[3p,times,procedia,sort&compress]{elsarticle}
\usepackage{nupha_ecrc}

\usepackage[usenames,dvipsnames]{color}
\definecolor{darkblue}{RGB}{0,0,196}
\usepackage[colorlinks=true,linkcolor=darkblue,citecolor=darkblue,urlcolor=darkblue]{hyperref}

\volume{00}

\firstpage{1}

\journalname{Nuclear Physics A}

\runauth{M. Strickland, M. Nopoush, and R. Ryblewski}

\jid{nupha}

\jnltitlelogo{Nuclear Physics A}

\usepackage{amssymb}

\usepackage[figuresright]{rotating}

\def\be{\begin{equation}}
\def\ee{\end{equation}}
\def\ba{\begin{eqnarray}}
\def\ea{\end{eqnarray}}

\begin{document}

\begin{frontmatter}



\dochead{}

\title{Anisotropic hydrodynamics for conformal Gubser flow}


\author[a,1]{Michael Strickland}
\author[a]{Mohammad Nopoush}
\author[b]{Radoslaw Ryblewski}
\address[a]{Kent State University, Kent OH 44242 USA}
\address[b]{The H. Niewodnicza\'nski Institute of Nuclear Physics, Polish Academy of Sciences, PL-31342 Krak\'ow, Poland}

\footnote{Presenter.}

\address{}

\begin{abstract}
In this proceedings contribution, we review the exact solution of the anisotropic hydrodynamics equations for a system subject to Gubser flow.  For this purpose, we use the leading-order anisotropic hydrodynamics equations which assume that the distribution function is ellipsoidally symmetric in local-rest-frame momentum.  We then prove that the SO(3)$_q$ symmetry in de Sitter space constrains the anisotropy tensor to be of spheroidal form with only one independent anisotropy parameter remaining.  As a consequence, the exact solution reduces to the problem of solving two coupled non-linear differential equations.  We show that, in the limit that the relaxation time goes to zero, one obtains Gubser's ideal hydrodynamic solution and, in the limit that the relaxation time goes to infinity, one obtains the exact free streaming solution obtained originally by Denicol et al.  For finite relaxation time, we solve the equations numerically and compare to the exact solution of the relaxation-time-approximation Boltzmann equation subject to Gubser flow.  Using this as our standard, we find that anisotropic hydrodynamics describes the spatio-temporal evolution of the system better than all currently known dissipative hydrodynamics approaches.
\end{abstract}

\begin{keyword}
anisotropic hydrodynamics, viscous hydrodynamics, kinetic theory, Gubser flow, exact solution
\end{keyword}

\end{frontmatter}


\section{Introduction}
\label{sec:intro}

Recently, there has been interest in obtaining exact solutions of anisotropic hydrodynamics \cite{Nopoush:2014qba}, viscous hydrodynamics \cite{Marrochio:2013wla,Hatta:2014gga,Hatta:2014gqa,Hatta:2014upa,Hatta:2014jva}, and kinetic theory \cite{Denicol:2014xca,Denicol:2014tha} subject to Gubser flow \cite{Gubser:2010ui,Gubser:2010ze}.  The reason for this interest is threefold:  (1) the exact solutions provide the possibility to obtain analytic expressions for the temperature evolution, various flow coefficients, etc., without the use of a complicated hydrodynamics simulations, (2) the exact solutions provide a way to test current hydrodynamic codes by setting up an initial condition corresponding to Gubser flow, and (3) the exact solutions provide a way to test the various anisotropic and viscous hydrodynamical approaches on the market.  The first reason is quite compelling, however, the phenomenological applicability of this type of flow is limited since it is conformal and corresponds to a system that has a quite strong transverse expansion at late times, irrespective of the coupling.  Despite that, this line of enquiry is quite interesting.  In this proceedings contribution, however, we will focus on the last reason for studying systems subject to Gubser flow to test different hydrodynamics approaches by comparing their predictions to a recently obtained exact solution of the Boltzmann equation in relaxation-time approximation subject to Gubser flow \cite{Denicol:2014xca,Denicol:2014tha}.

In the context of relativistic heavy-ion collisions, the use of dissipative viscous hydrodynamics is now common (see e.g. Refs.~\cite{Romatschke:2009im,Gale:2013da,Jeon:2015dfa} and references therein), however, traditional viscous hydrodynamics approaches rely on linearization around an isotropic equilibrium state.  If the system has large non-equilibrium corrections, a perturbative treatment may not be phenomenologically reliable at all points in spacetime.  In order to address this issue, the framework of anisotropic hydrodynamics \cite{Martinez:2010sc,Florkowski:2010cf} was created in order to extend the range of applicability of dissipative hydrodynamics (see Ref.~\cite{Strickland:2014pga} for a recent review).  In the anisotropic hydrodynamics framework, the most important (diagonal) components of the energy-momentum tensor are treated non-perturbatively and non-spheroidal/off-diagonal components are treated perturbatively.  This approach has been shown to more accurately describe the evolution of systems subject to boost-invariant and transversely homogeneous (0+1)-dimensional flow than traditional viscous hydrodynamics approaches \cite{Florkowski:2013lza,Florkowski:2013lya,Bazow:2013ifa,Nopoush:2014pfa,Bazow:2015cha}.  Here we report on the work presented in Ref.~\cite{Nopoush:2014qba}, in which the exact solution of the leading-order anisotropic hydrodynamics equations subject to Gubser flow was obtained.

\section{Gubser flow}

Herein, we assume that the system is boost invariant and cylindrically symmetric with respect to the beam line at all times.  With this assumption, one can construct a flow with SO(3)$_q\,{\otimes}\,$SO(1,1)$\,{\otimes}\,$Z$_2$ symmetry~\cite{Gubser:2010ui,Gubser:2010ze}.  In this case, all dynamical variables depend on $\tau = \sqrt{t^2-z^2}$ and $r = \sqrt{x^2+y^2}$ through the dimensionless combination $G(\tau,r)=(1-q^2{\tau}^2+q^2r^2)/(2 q{\tau})$, where $q$ is an arbitrary energy scale which sets the physical size of the system.  The resulting flow is completely determined by symmetry constraints to be $\tilde{u}^\mu = (\cosh\theta_\perp,\sinh\theta_\perp,0,0)$, with $\tanh\theta_\perp \equiv 2q^2\tau r/(1+q^2\tau^2+q^2r^2)$ and the tilde indicating polar Milne coordinates with position four-vector $\tilde{x}^\mu=(\tau,\,r,\,\phi,\,\varsigma)$ and $\phi = \tan^{-1}(y/x)$ and $\varsigma = \tanh^{-1}(z/t)$.  

To map this to a static flow, one performs a Weyl rescaling and a change of variables to de Sitter coordinates $\sinh{\rho} =  - (1-q^2{\tau^2}+q^2r^2)/(2q{\tau})$ and $\tan{\theta} = 2qr/(1+q^2{\tau}^2-q^2r^2)$, where $\rho$ is interpreted as the de Sitter ``time'' and $\theta$ is an angular variable.  Due to the symmetry of the flow, physical quantities only depend on $\rho$.  For fixed $r$, the limit $\tau \rightarrow 0^+$ corresponds to the limit $\rho \rightarrow -\infty$ and the limit $\tau\rightarrow\infty$ corresponds to the limit $\rho \rightarrow \infty$.  This means that the de Sitter map covers the future (forward) light cone.  Finally, we note that in the remainder of the paper, Weyl-rescaled de Sitter-space quantities are indicated with a hat, e.g. the position four-vector becomes $\hat{x}^\mu=(\rho,\,\theta,\,\phi,\,\varsigma)$.

\section{Exact anisotropic hydrodynamics equations}

We now introduce our ansatz for the one-particle distribution function.  Since the system is cylindrically symmetric with respect to the beam line, the de Sitter space anisotropy tensor can be assumed to be diagonal.\footnote{Any off-diagonal contributions will quickly relax to zero if they are not assumed to be zero.}   An ellipsoidal anisotropic distribution function can be constructed by introducing a tensor of the form $\hat{\Xi}^{\mu\nu}=\hat{u}^\mu \hat{u}^\nu+\hat{\xi}^{\mu\nu}$, where $\hat{u}^\mu$ is the four-velocity and $\hat{\xi}^{\mu\nu}$ is a symmetric traceless anisotropy tensor \cite{Nopoush:2014pfa}.  Expanding $\hat{\xi}^{\mu\nu}$ in the de Sitter-space basis gives $\hat{\xi}^{\mu\nu} = \hat{\xi}_\theta \hat{\Theta}^\mu \hat{\Theta}^\nu+\hat{\xi}_\phi \hat{\Phi}^\mu \hat{\Phi}^\nu+\hat{\xi}_\varsigma \hat{\varsigma}^\mu \hat{\varsigma}^\nu$, where $\hat{\Theta}^\mu$, $\hat{\Phi}^\mu$, and $\hat{\varsigma}^\nu$ are Weyl-rescaled de Sitter space basis vectors which obey $\hat{u}^\mu \hat{u}_\mu =-1$, $\hat{\Theta}^\mu\hat{\Theta}_\mu=1$, $\hat{\Phi}^\mu\hat{\Phi}_\mu=1$, $\hat{\varsigma}^\mu\hat{\varsigma}_\mu=1$.\footnote{Our Minkowski-space metric convention is $g^{\mu\nu} = {\rm diag}(-1,1,1,1).$}   The anisotropy tensor is traceless, i.e. $\hat{\xi}^{\mu}_{\ \mu} = 0$, and orthogonal to the flow, i.e. $\hat{u}_\mu \hat{\xi}^{\mu\nu} = 0$.  Using the tensor $\hat{\Xi}^{\mu\nu}$, one can construct an anisotropic distribution function for a conformal system~\cite{Nopoush:2014pfa}
\be
f(\hat{x},\hat{p})=f_{\rm iso}\left(\frac{1}{\hat\lambda}\sqrt{\hat{p}_\mu\hat{\Xi}^{\mu\nu} \hat{p}_\nu}\right)\, ,
\label{eq:pdf}
\ee
where we have assumed vanishing chemical potential and $\hat\lambda$ is non-equilibrium scale (transverse temperature) which can be identified with the de Sitter-space temperature, $\hat{T}$, only when $\hat\xi^{\mu\nu}=0$.

To determine the $\rho$-dependence of the scale $\hat\lambda$ and anisotropies $\hat\xi_i$, we take moments of the Boltzmann equation in relaxation-time approximation $\hat{p} \cdot D f = \hat{p}\cdot \hat{u} \, (f-f_{\rm iso})/\hat\tau_{\rm eq}$, where $D_\mu$ is the covariant derivative, $f_{\rm iso}$ denotes the isotropic equilibrium distribution function, and $\hat\tau_{\rm eq}$ is the relaxation time. For a conformal system in relaxation-time approximation, one has $\hat\tau_{\rm eq} = 5\hat{\bar\eta}/\hat{T}$, where $\hat{\bar\eta}=\hat\eta/\hat{s}=\eta/s$ with $\hat\eta$ and $\hat{s}$ being the Weyl-rescaled shear viscosity and entropy density, respectively.  Using the first and second moments of the Boltzmann equation in de Sitter coordinates, one obtains two coupled ordinary differential equations~\cite{Nopoush:2014qba}
\ba
4\frac{d\log\hat\lambda}{d\rho}+\frac{3 \hat\alpha_\varsigma^2\left(\frac{H_{2
   L}(\bar{y})}{H_2(\bar{y})}+1\right)-4}{3\hat\alpha_\varsigma^2-1} \, \frac{d\log\hat\alpha_\varsigma}{d\rho}+ \tanh\rho\left(\frac{H_{2T}(\bar{y})}{H_2(\bar{y})}+2\right) &=& 0\, ,
\label{eq:1st-mom-final2} \\
\frac{6\hat\alpha_{\varsigma }}{1-3 \hat\alpha _\varsigma ^2} \frac{d \hat\alpha_\varsigma}{d\rho} -\frac{3 \left(3 \hat\alpha_\varsigma^4-4\hat\alpha_\varsigma^2+1\right)}{4\hat\tau_{\rm eq} \hat\alpha _{\varsigma }^5} \left(\frac{\hat{T}}{\hat\lambda}\right)^5+2\tanh\rho=0 \, ,
\label{eq:2nd-mom-final2}
\ea
where $\hat\alpha_i \equiv (1+\hat{\xi}_i)^{-1/2}$, $\bar{y}^2 \equiv (3\hat\alpha_\varsigma^2-1)/2$, and $\hat{T}=\hat\alpha_\varsigma \hat\lambda \left(H_2(\bar{y})/2\right)^{1/4}\!/\bar{y}$. The definitions of the $H$-functions appearing above can be found in Ref.~\cite{Nopoush:2014qba}.

\begin{figure}[t]
\hspace{-4mm}\includegraphics[width=1.02\linewidth]{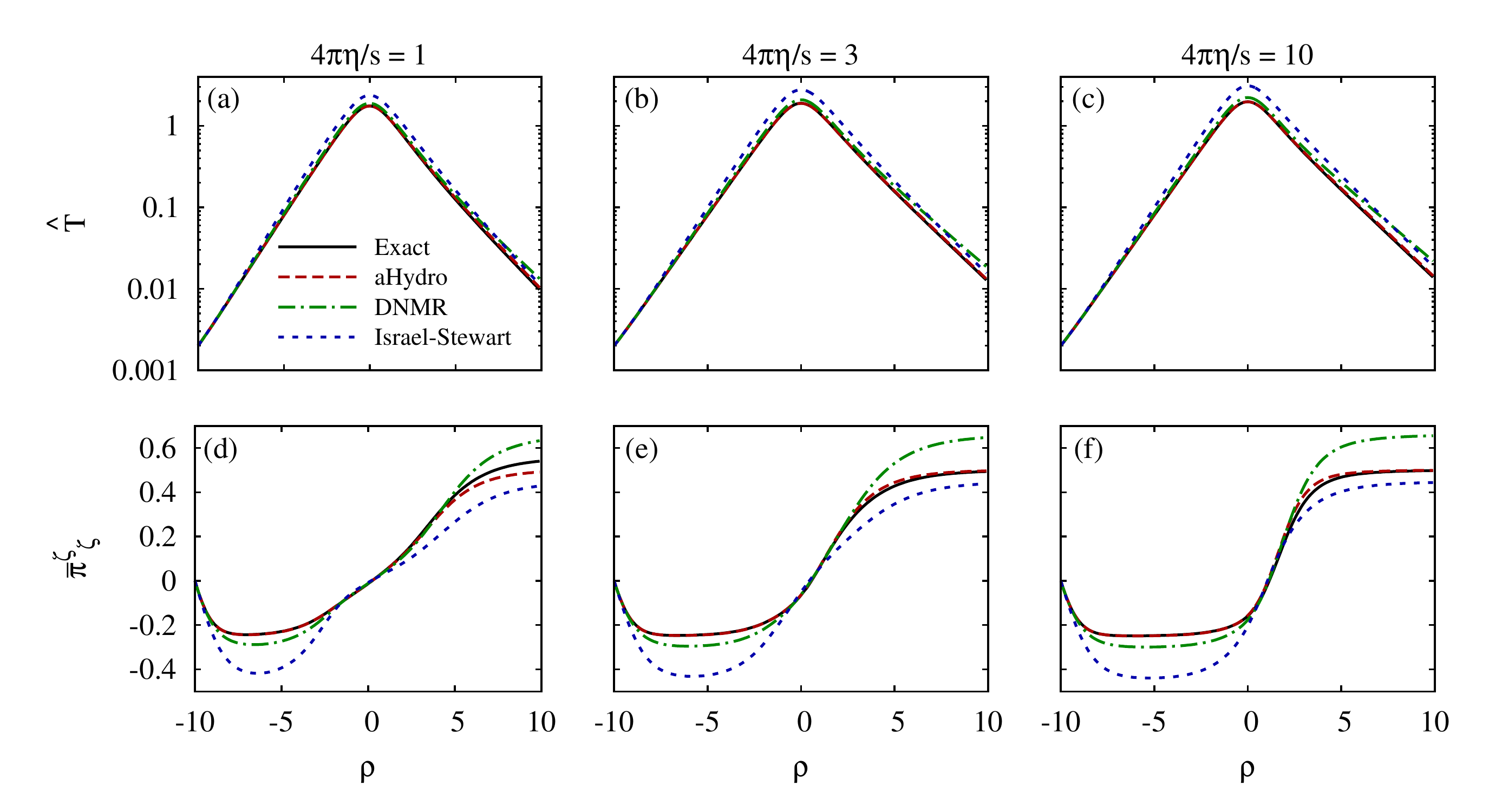}
\vspace{-4mm}
\caption{In the top row, we compare the de Sitter-space effective temperature $\hat{T}$ obtained from the exact kinetic solution obtained in Refs.~\cite{Denicol:2014xca,Denicol:2014tha} (black solid line), the leading-order anisotropic hydrodynamics equations obtained in Ref.~\cite{Nopoush:2014qba} (red dashed line), DNMR second-order viscous hydrodynamics obtained in Ref.~\cite{Marrochio:2013wla} (green dot-dashed line), and Israel-Stewart second-order viscous hydrodynamics obtained in Ref.~\cite{Marrochio:2013wla} (blue dotted line).  The columns from left to right correspond to three different choices of the shear viscosity to entropy density ratio with $4 \pi \eta/s \in \{1,3,10\}$, respectively.  In the bottom row, we compare results for the scaled shear $\bar{\pi}_{\varsigma}^{\varsigma} \equiv \hat{\pi}_\varsigma^\varsigma/(\hat{T}\hat{s})$.  The labeling and values of $4 \pi \eta/s$ in the bottom row are the same as in the top row.  In all cases, at $\rho=\rho_0=-10$, we fixed the initial effective temperature to be $\hat{T}_0 = 0.002$ and the initial anisotropy to be  $\hat\alpha_{\varsigma,0}=1$, which corresponds to an isotropic initial condition in de Sitter space.
}
\label{fig:fig-rho-10-T0p002-az1}
\end{figure}

\section{Results and conclusions}

Using Eqs.~(\ref{eq:1st-mom-final2}) and (\ref{eq:2nd-mom-final2}) it was shown in Ref.~\cite{Nopoush:2014qba} that, in the limit that the relaxation time goes to zero, one obtains Gubser's exact ideal hydrodynamic solution and, in the limit that the relaxation time goes to infinity, one obtains the exact free streaming solution obtained originally by Denicol et al.~\cite{Denicol:2014tha}.  For finite relaxation time, one can solve Eqs.~(\ref{eq:1st-mom-final2}) and (\ref{eq:2nd-mom-final2}) numerically.  As usual, in order to complete the solution, one needs to specify a boundary condition.  Since we want to have a solution in the entire forward lightcone, we fix the boundary condition on the ``left'' ($\rho_0 \rightarrow -\infty$), which maps to $\tau \rightarrow 0^+$.  Note that, with this boundary condition, one can smoothly take the limit $\eta/s \rightarrow 0$ in order to obtain Gubser's exact ideal hydrodynamic solution (see Fig. 8 of Ref.~\cite{Denicol:2014tha}).  This limit is not guaranteed for other choices of $\rho_0$ \cite{Heinz:2015cda}.

As can be seen from Fig.~\ref{fig:fig-rho-10-T0p002-az1}, the anisotropic hydrodynamics equations obtained in Ref.~\cite{Nopoush:2014qba} provide the best approximation to the exact result.  For the temperature, it is very difficult to distinguish the anisotropic hydrodynamics result from the exact result.  For the scaled shear $\bar{\pi}_{\varsigma}^{\varsigma} \equiv \hat{\pi}_\varsigma^\varsigma/(\hat{T}\hat{s})$, there are visible differences between the anisotropic hydrodynamics solutions and the exact solution in the region above $\rho \gtrsim  0$.  In all cases, at large $\rho$ one sees that anisotropic hydrodynamics has the correct asymptotic behavior.  The latter observation can be proven analytically~\cite{Nopoush:2014qba}.  In Ref.~\cite{Nopoush:2014qba} anisotropic initial conditions were also considered with the conclusion being the same.  Based on these findings we conclude that anisotropic hydrodynamics describes the spatio-temporal evolution of the system better than all currently known dissipative hydrodynamics approaches.

Note that the solutions reviewed herein can be easily mapped back to Milne space, giving the full spatio-temporal evolution for a boost-invariant and cylindrically-symmetric system for arbitrary values of parameter $q$ in the entire forward lightcone.  Using this mapping, one can obtain the radial temperature profile at any given proper time for an arbitrary system size set by the scale $q$.  This can be used as an initial condition for subsequent evolution in Milne space.  In Ref.~\cite{Nopoush:2015yga} comparison of the solutions of (1+1)-dimensional ellipsoidal anisotropic hydrodynamics equations of Tinti and Florkowski \cite{Tinti:2013vba} with the exact solution obtained herein demonstrated that the framework of Tinti and Florkowski was able to numerically reproduce the exact solution without having to transform to de Sitter space.   The results presented here can now be used as a test case for code validation in all future anisotropic hydrodynamics codes.

\section*{Acknowledgments}

We thank G.~Denicol, M.~Martinez, U.~Heinz, and J.~Noronha for useful discussions.  M. Strickland and M. Nopoush were supported by the U.S. Department of Energy, Office of Science, Office of Nuclear Physics under Award No.~DE-SC0013470.  R.R. was supported by Polish National Science Center Grant No.~DEC-2012/07/D/ST2/02125 and the Polish Ministry of Science and Higher Education fellowship for young scientists (X edition).





\bibliographystyle{elsarticle-num}
\bibliography{gubserQM}







\end{document}